# The Bright Symbiotic Mira EF Aquilae

Bruce Margon, J. Xavier Prochaska, and Nicolas Tejos

Department of Astronomy & Astrophysics, and University of California Observatories,

University of California, Santa Cruz, 1156 High Street, Santa Cruz, CA 95064;

margon@ucsc.edu; xavier@ucolick.org; ntejos@gmail.com

and

TalaWanda Monroe

Space Telescope Science Institute, 3700 San Martin Drive, Baltimore, MD 21218

tmonroe@stsci.edu

**ABSTRACT**. An incidental spectrum of the poorly studied long period variable EF Aquilae shows [O III] emission indicative of a symbiotic star. Strong GALEX detections in the UV reinforce this classification, providing overt evidence for the presence of the hot subluminous companion. Recent compilations of the photometric behavior strongly suggest that the cool component is a Mira variable. Thus EF Aql appears to be a member of the rare symbiotic Mira subgroup.

The long period variable EF Aquilae, although noted as a variable star more than a century ago (Reinmuth 1925, who cites observations as early as 1902), is poorly studied. The General Catalogue of Variable Stars (Samus et al. 2007) suggests $12.4 < V < 15.5$. It appears as a very bright IRAS, 2MASS, and WISE source, but we are unable to locate any published spectroscopic information.

Several of the modern synoptic photometric surveys contain coverage of EF Aql, including ASAS (Pojmanski 2002), NSVS (Wozniak et al. 2004), and CRTS (Drake et al. 2014). Long term photometric variations are clearly present in all of these data sets, with most of the data bases suggesting a period of near 320 days, or half that, and amplitudes of 2-3 mag in the visual band. Further refinement of the period and light curve may be somewhat awkward due to the close proximity of the suggested period to one year. Richwine et al. (2005) have examined the synoptic survey data for EF Aql and recommend a period of 329.4 d, with amplitude >2.4 mag, and classify the object as a Mira. LeBertre et al. (2003) provide K and L' photometry for EF Aql, and classify it as O-rich.

As part of a program to identify QSOs with near-ultraviolet excesses (the UV bright Quasar Survey; UVQS; Monroe et al. 2015), we obtained an (obviously incidental) spectrum of EF Aql on UT 2014 August 2, using the 2.5m du Pont telescope of the Las Campanas Observatory. A 180 s exposure with the 600/5000 grating and 2" slit yielded ~4 Å spectral resolution. Approximate flux calibration was obtained via observations of spectrophometric standard stars. The resulting spectrum appears in Figure 1.

The spectrum is typical of a classical LPV, with multiple strong TiO bands and prominent Balmer emission visible through at least H11, and possibly to H17. The high TiO opacity suppresses H$\beta$ and H$\gamma$, leaving a very strong H$\delta$ line (Joy 1926, 1947). H$\epsilon$ is overwhelmed by the strong Ca II H-line absorption. Of particular interest is the clear detection of [O III] $\lambda$5007 emission. The expected, weaker $\lambda$4959 line is likely marred by the adjacent TiO $\lambda$4960 bandhead. It therefore appears that EF Aql is a symbiotic star, having thus far escaped notice due to the lack of a published spectrum.

Our spectrum is indeed reminiscent of that of the well-studied symbiotic Mira R Aqr near maximum light presented by Munari & Zwitter (2002, Figure 98), which also displays prominent TiO absorption and H$\delta$ emission, and weak but distinct [O III] $\lambda$5007 emission, with [O III] $\lambda$4959 obscured by TiO. Although the absolute flux calibration of our EF Aql spectrum is crude (due to slit losses and non-photometric conditions), it is consistent with the system also being at maximum, near V~12, strengthening the spectral resemblance to R Aqr. The photometric ephemeris for AE Aql listed by the AAVSO is also consistent with our spectrum occurring near maximum light, although given the considerable uncertainties, this agreement may well be fortuitous. HeI $\lambda$5876 emission is probably also present in our spectrum, as in the symbiotic RR Tel (McKenna et al. 1997), but this requires confirmation. The $\lambda$6825 Å Raman scattered emission line characteristic of symbiotics (Webster & Allen 1975, Bond & Kasliwal 2012) is unfortunately just off the red end of our spectral coverage.

Aside from the photometric and spectroscopic observations noted above, the IR colors of this object also confirm the likely symbiotic Mira classification. Allen (1982) pointed out that (J-H), (H-K) colors of symbiotics segregate these systems into two groups. EF Aql, with 2MASS colors of (J-H)=0.95, (H-K)= 0.66, falls comfortably into Allen's "D" (dusty) class, and Whitelock (1987) stresses that these "D" symbiotic systems have been found to be Miras.

We may use the well-known period luminosity relations for Miras to estimate a distance to EF Aql. The maps of Schlegel et al. (1998) suggest $A_V = 0.45$ at this moderate galactic latitude (b= -16°), so extinction at K-band should be negligible. If we employ the 329 d period preferred by Richwine et al. (2005) and the 2MASS K = 5.36 magnitude, then the Gromadzki et al. (2009) correlations for symbiotic Miras imply $M_K = -7.7$, in agreement with many estimates for these stars. As the relation is logarithmic in the period, the period uncertainty is not a major factor. The implied distance is 4.2 kpc.

The symbiotic classification is further supported by the detection of this star by GALEX, with NUV = 16.01 (equivalent to a bandpass of 1750-2800 Å), the factor that led our attention to this system originally. The anomalously bright NUV flux from this star does not appear to have been previously noted, and is no doubt an overt detection of the hot subluminous companion present in symbiotic systems, as the M primary contributes negligible flux in this bandpass. Inferences on the nature of this companion are necessarily uncertain for multiple reasons. Although extinction at K can be neglected, extinction in the GALEX NUV band certainly cannot. The Bianchi et al. (2005) relations suggest $A_{NUV} \sim 8$ E(B-V), requiring a large (at least

~3×) and uncertain correction to the observed GALEX flux. Further, with only 2 GALEX bands, a proper model atmosphere fit is clearly not practical. Nonetheless, at the 4.2 kpc distance inferred above, even crude estimates show that the hot source is likely more luminous than a white dwarf, and thus may well be a subdwarf.

Whitelock (2003) remarks that symbiotic Miras are quite rare. Although a few such systems have been intensively studied, e.g., (R Aqr, RR Tel, and of course the eponymous Mira), there are probably less than two dozen currently cataloged systems (Whitelock 1987). Thus the addition of EF Aql to the list is of some interest. Brighter than many of the previously known symbiotic Miras, and at an equatorial location, EF Aql should be useful in further study of this interesting stellar subset.

We are grateful to the anonymous referee, Karen Kwitter, and especially Howard Bond for many useful discussions.

**FIGURE LEGEND**

Figure 1. The spectrum of EF Aql. The [O III] λ5007 emission is evident, as well as the suppression of Hβ and Hγ emission caused by high TiO opacity.

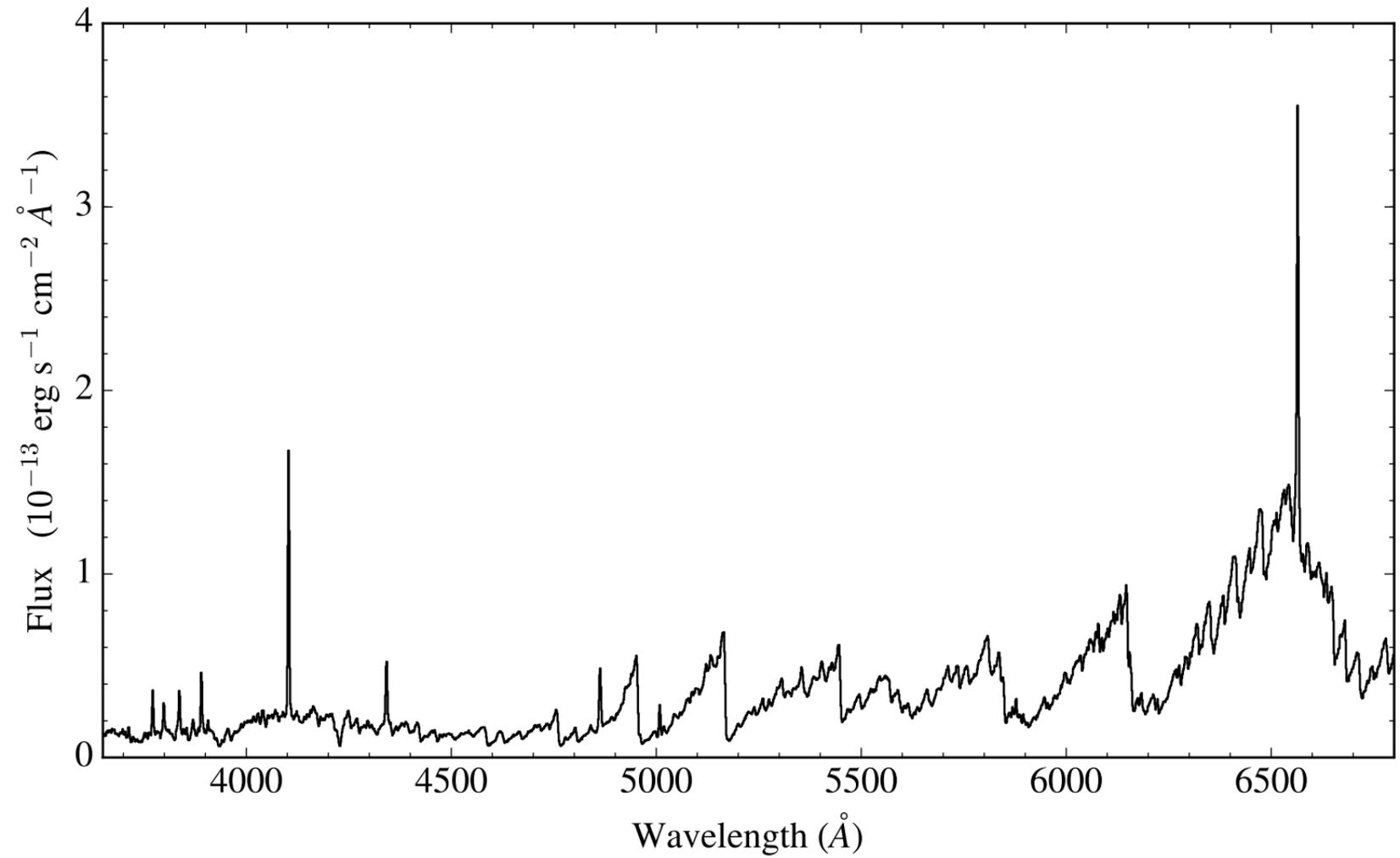